\documentclass[onecolumn,a4]{article} 
\usepackage{version}
\usepackage{graphicx}
\usepackage[]{natbib}
\usepackage{txfonts}
\usepackage{booktabs}
\usepackage{multirow}
\newcommand{\ra}[1]{\renewcommand{\arraystretch}{#1}}
\def\lsim{\mathrel{\rlap{\lower4pt\hbox{\hskip1pt$\sim$}}
    \raise1pt\hbox{$<$}}}                
\def\gsim{\mathrel{\rlap{\lower4pt\hbox{\hskip1pt$\sim$}}
    \raise1pt\hbox{$>$}}}                
\usepackage{draftwatermark}
\SetWatermarkText{Accepted in A\&A}
\SetWatermarkScale{0.75}
\SetWatermarkAngle{60}

\usepackage{rotating}
\usepackage{pdflscape}
\newcommand{\jcp}{Journal of Chem. Phys.}

\newcommand{\apj}{ApJ}
\newcommand{\apjl}{ApJ}
\newcommand{\apjs}{ApJS}

\newcommand{\aap}{A\&A}
\newcommand{\aj}{AJ}

\newcommand{\mnras}{MNRAS}
\newcommand{\aaps}{A\&AS}
\newcommand{\icarus}{Icarus}
\newcommand{\ssr}{Space~Sci.~Rev.}

\begin{document}
%
  \section*{Heavy ion irradiation of crystalline water ice} 
  \subsection*{Cosmic ray amorphisation cross-section and sputtering yield}

   E. Dartois  $^{1,2}$
          \footnote{Experiments performed at the Grand Acc\'el\'erateur National d'Ions Lourds (GANIL) Caen, France. Part of this work has been financed by the French INSU-CNRS programme ``Physique et Chimie  du Milieu Interstellaire'' PCMI) and the ANR IGLIAS},
          B.~Aug\'e $^{3,1,2}$,
          P.~Boduch $^{3}$,
          R.~Brunetto $^{1,2}$,
          M.~Chabot $^{4,5}$,
          A.~Domaracka $^{3}$,
          J. J. Ding $^{3}$,
          O.~Kamalou $^{6}$,
         X.~Y.Lv$^{3}$,
          H.~Rothard $^{3}$,
         E.~F.~da~Silveira$^{7}$,
          J.~C.~Thomas $^{6}$\\

\noindent $^{1}$ CNRS-INSU, Institut d'Astrophysique Spatiale, UMR 8617, 91405 Orsay, France\\
$^{2}$ Universit\'e Paris Sud, Institut d'Astrophysique Spatiale, UMR 8617, b\^atiment 121, 91405 Orsay, France\\
$^{3}$ Centre de Recherche sur les Ions, les Mat\'eriaux et la Photonique (CEA/CNRS/ENSICAEN/Universit\'e de Caen-Basse Normandie),
UMR~6252, CIMAP - CIRIL - Ganil, Boulevard Henri Becquerel, BP 5133, 14070 Caen Cedex 05, France\\
$^{4}$ CNRS-IN2P3, Institut de Physique Nucl\'eaire d'Orsay, UMR8608, 91406 Orsay, France\\
$^{5}$ Universit\'e Paris Sud, Institut de Physique Nucl\'eaire d'Orsay, UMR8608, IN2P3-CNRS, b\^atiment 103, 91406 Orsay, France\\
$^{6}$ Grand Acc\'el\'erateur National d'Ions Lourds, CEA/DSM-CNRS/IN2P3, Boulevard Henri Becquerel, BP 55027, 14076 Caen Cedex 05, France\\
$^{7}$ Departamento de F\'isica, Pontif\'\i cia Universidade Cat\'olica do Rio de Janeiro, Rua Marqu\^es de S\~ao Vicente 225, 22451-900, Rio de Janeiro, RJ, Brazil\\


 
  \section*{Abstract}
{Under cosmic irradiation, the interstellar water ice mantles evolve towards a compact amorphous state. Crystalline ice amorphisation was previously monitored mainly in the keV to hundreds of keV ion energies.}
{We experimentally investigate heavy ion irradiation amorphisation of crystalline ice, at high energies closer to true cosmic rays, and explore the water-ice sputtering yield.}
{We irradiated thin crystalline ice films with MeV to GeV swift ion beams, produced at the GANIL accelerator. The ice infrared spectral evolution as a function of fluence is monitored with in-situ infrared spectroscopy (induced amorphisation of the initial crystalline state into a compact amorphous phase).}
{The crystalline ice amorphisation cross-section is measured in the high electronic stopping-power range for different temperatures. At large fluence, the ice sputtering is measured on the infrared spectra, and the fitted sputtering-yield dependence, combined with previous measurements, is quadratic over three decades of electronic stopping power.}
{The final state of cosmic ray irradiation for porous amorphous and crystalline ice, as monitored by infrared spectroscopy, is the same, but with a large difference in cross-section, hence in time scale in an astrophysical context. The cosmic ray water-ice sputtering rates compete with the UV photodesorption yields reported in the literature. The prevalence of direct cosmic ray sputtering over cosmic-ray induced photons photodesorption may be particularly true for ices strongly bonded to the ice mantles surfaces, such as hydrogen-bonded ice structures or more generally the so-called polar ices.}
%

%

\section{Introduction}
\label{intro}

Crystalline ice is observed in the interstellar medium in the inner regions around young stellar objects 
\citep[e.g.][and articles referencing them]{Eiroa1983,Brooke1996,Dartois1998,Brooke1999,Molinari1999,Malfait1999,Gibb2004, Dartois2005,Boogert2008,Oberg2011} and in the warm dust grains around evolved late-type OH-IR stars
\citep[e.g.][]{Soifer1981, Roche1984, Meyer1998, Sylvester1999, Dartois2002, Maldoni2003, Justtanont2006, Lombaert2013, Suh2013}.\\
The interstellar ice mantles present in these environments are immersed in a flux of cosmic ray particles \citep[e.g.][]{Simpson1983, Webber1983} that produces new species via radiolytic processes, but first affects their structural changes \citep[][]{Strazzulla1992, Moore1992, Leto2003, Baragiola2005, Mastrapa2006, Fama2010, Dartois2013} and induces molecules and radicals desorption from these grains \citep[e.g.][]{Shen2004}.\\
Inside dense clouds, cosmic rays are the source of the internal FUV (photon energies above about 6 eV) photochemistry in addition to radiolysis.
Both the FUV photon field and cosmic rays colliding with dust grains are major sources of desorption, driving a gas to solid state equilibrium, preventing the rapid freeze-out of all the species in the solid phase and participating in the construction of molecular complexity.
The photodesorption and cosmic ray desorption rates, as well as  
the corresponding astrophysical time scales on which these processes are effective, can be derived with dedicated ion irradiations of ice films, and the magnitude of the effect can be
understood with such experimental facilities.\\
In this article we investigate the amorphisation of crystalline water ice irradiated by swift heavy ions. The experiments performed are described in Sect. 2 and the  results obtained explained in Sect. 3. These measurements are discussed in Sect. 4 in the framework of astrophysical implications. A conclusion is drawn in Sect. 5.

\section{Experiments}

Swift ion irradiation experiments were performed at the heavy-ion accelerator Grand Acc\'el\'erateur National d'Ions Lourds (GANIL, Caen, France).
Heavy ion projectiles were delivered on the IRRSUD, SME, and LISE beam lines\footnote{http://pro.ganil-spiral2.eu/laboratory/experimental-areas} between July 2008 and November 2014. These beams were coupled to the CASIMIR (Chambre d'Analyse par Spectroscopie Infrarouge des Mol\'ecules IRradi\'ees) facility, a high vacuum chamber (a few 10$^{-8}$ mbar) holding an infrared transmitting substrate that can be cryocooled down to about 13.6 K. The ice films are produced by placing the cold window substrate in front of a deposition line. (Poly-)crystalline ice films were condensed at 150~K on the window, from the vapour phase, then cooled down to the lowest temperature of the cold finger and returned to the temperature set for irradiation. We refer to this initial state as "crystalline ice" within the article. The water ice sample, whose optical depth spectra as a function of fluence are shown in Fig. \ref{Ni_77}, experienced a different deposition history. The ice mantle has been formed by annealing of a film deposited at low temperature (13.6~K) of H$_2$O co-mixed with N$_2$ (and traces of CH$_4$) from a previous set of experiments. The resulting (poly-)crystalline structure after annealing at 150K - the N$_2$ component sublimates but not H$_2$O -  and back to 77K is thus slightly different from the other films, and is analysed in the same way.
Details of the experimental setup are given in \cite{Melot2003} and \cite{Seperuelo2009}.
The film thickness is chosen to give a high band contrast for the infrared absorptions, without saturating the bands. (The 40~K experiment infrared signals are close to saturation.) In addition, the chosen thicknesses allow the ion beam to pass through the film with an almost constant energy loss per unit path length.
A Nicolet FTIR spectrometer (Magna 550) with a spectral resolution of 1 cm$^{-1}$ was used to monitor the infrared film transmittance. 
The chamber is equipped with a cryostat that can rotate up to 180 degrees, and the irradiation and recording of the transmittance spectra were performed at normal incidence.
The projectiles energies were chosen to cover the high stopping-power range, to complement existing data at lower energies with lighter ions.
The temperatures were selected to explore the amorphisation dependencies discussed in the literature for lower projectiles energies or electrons.
The evolution of the spectra was recorded at several fluences.

\section{Results}
\label{results}
A summary of the experiments performed and considered in the forthcoming discussion is given in Table~\ref{summary}. The column density evolution of the water ice molecules submitted to ion irradiation can be described, to first order, as a function of ion fluence (F), by a coupled set of differential equations:
\begin{equation} 
\left\lbrace
\begin{array}{ccl}
\rm dN/dF       &\rm =  &\rm -\sigma_d N - Y_s \times f\\
\rm dw_c/dF     &\rm =  &\rm -\sigma_{am} w_c
\end{array}\right.
\end{equation}
where N is the total H$_2$O column density, $\rm \sigma_d$ their effective radiolytic destruction cross-section (cm$^{2}$), $\rm Y_s$ the H$_2$O sputtering yield (H$_2$O/ion), multiplied, to first order, by the relative concentration f of water ice molecules with respect to the total number of molecules/radicals in the ice film. Here, $\rm \sigma_{am}$ is the crystalline ice amorphisation cross-section, and $\rm w_c$ the crystalline fraction of the ice. 
The $\rm -\sigma_d N$ and $\rm Y_s \times f$ terms in the top part of eq.1 are related to the destruction and sputtering contributions to the evolution of the ice column density.
Pure water ice is very radiolytically resistant. One of the main products formed by irradiation is H$_2$O$_2$, which saturates to a few percent of water-ice molecules, by number, at doses above about 10 eV/molecule (see e.g. \citealt{Moore2000} for the production yield of H$_2$O$_2$). The same low efficiency is observed for FUV irradiation of pure ice by \cite{Gerakines1996}, among others.

\begin{figure}[htbp]
\begin{center}
\includegraphics[height=\columnwidth,angle=90]{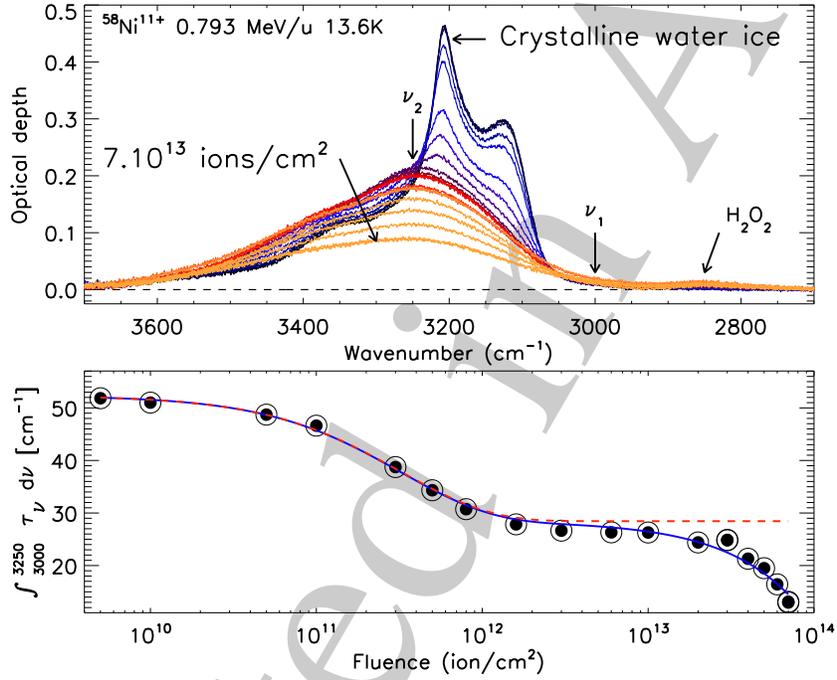}
\caption{Upper panel: optical depth spectrum evolution upon 46~MeV Ni swift ion irradiation at 13.6~K of a thick water ice film crystallised at 150~K. The crystalline ice spectrum and the spectrum at the highest fluence are labelled; the intermediate spectra correspond to the fluence evolution given in the lower panel.
Lower panel: Integrated optical depth $\rm I_{OD}$ over the ($\rm\nu_1=$) 3000 to ($\rm\nu_2=$) 3250~cm$^{-1}$ water-ice stretching-mode profile interval, where the amorphous-to-crystalline contrast is higher, as a function of fluence (filled circles). A model fit (from Eq. 5) is overplotted (full lines). The fit obtained without any water ice sputtering contribution is also shown (dashed lines) for comparison.}
\label{Ni_13_6}
\end{center}
\end{figure}

Considering the radiolytic resilience of pure water ice, we impose that f remains close to unity. For our considered experimental fluences and ice film thicknesses, if the radiolysis is negligible, then the evolution of the water ice spectra is dominated by the crystalline to compact amorphous phase change and sputtering. The solutions to these equations come down to
%
\begin{equation} 
\left\lbrace
\begin{array}{ccl}
\rm N   &\rm \approx    &\rm N_0 - Y_s\,F\\
\rm w_c         &\rm =  &\rm exp(-\sigma_{am} F) = N_c/(N_c+N_a) = N_c / N
\end{array}\right.
\end{equation}
where N$\rm_c$ and N$\rm_a$ are the crystalline and amorphous ice column densities, respectively.
The column densities of the molecules are followed experimentally in the infrared through
\begin{equation} 
\rm N = \int^{\bar{\nu}2}_{\bar{\nu}1} \tau(\bar{\nu})d\bar{\nu}/A \, ,
\end{equation}
where the integral of optical depth ($\rm \tau$) is taken over the band frequency range, and A is the band strength value (cm/molecule) for the vibrational mode and ice structure (compact amorphous or crystalline) considered.  During the irradiation, the phase change from a crystalline to an amorphous compact state induces a variation in the band strengths \citep[see e.g. ][]{Leto2003}. Therefore, assuming a direct impact model,
\begin{equation} 
\rm N = \frac{1}{A_c}\int^{\bar{\nu}2}_{\bar{\nu}1} \tau(\bar{\nu})d\bar{\nu} \; [w_c+ \alpha(1-w_c)] ; \alpha=A_c/A_{am} ,
\end{equation}
where $\rm A_c$ and $\rm A_{am}$ are the band strength values for crystalline and compact amorphous ice (integrated over the same wavenumber range).
The measured integrated optical depth, $\rm I_{OD} = \int^{\bar{\nu}2}_{\bar{\nu}1} \tau(\bar{\nu})d\bar{\nu}$, can then be fitted with
%
%
\begin{equation} 
\rm I_{OD} \approx \frac{A_c}{w_c+ \alpha(1-w_c)} \left[ N_0 - Y_s F\right]
.\end{equation}
To optimise the crystalline to amorphous transition cross section determination, the analysis is performed on the spectral region where the contrast between amorphous and crystalline ice absorption profiles is maximum. The band strength value variations expected over the full profile between crystalline, low-temperature porous amorphous and compact amorphous phases are in the several tenths of a percent range \citep[see e.g.][]{Leto2003}. 
However, the most intense profiles changes lie in the range covering the specific crystalline ice peaks, around 3130 and 3210~cm$^{-1}$ \cite[see e.g. Fig.~1 in][and Fig.~\ref{Ni_13_6} in this article]{Leto2003}. We thus perform the spectral analysis over the 3000 to 3250~cm$^{-1}$ profile interval (shown in Fig~\ref{Ni_13_6}), where the contrast between crystalline and compact amorphous band strength value variations is maximum.
Similar contrast approaches are often considered by authors using the near-infrared 1.65 $\mu$m overtone mode \citep[e.g.][]{Leto2005,Mastrapa2006, Fama2010}.

\begin{figure}[htbp]
\begin{center}
\includegraphics[height=\columnwidth,angle=90]{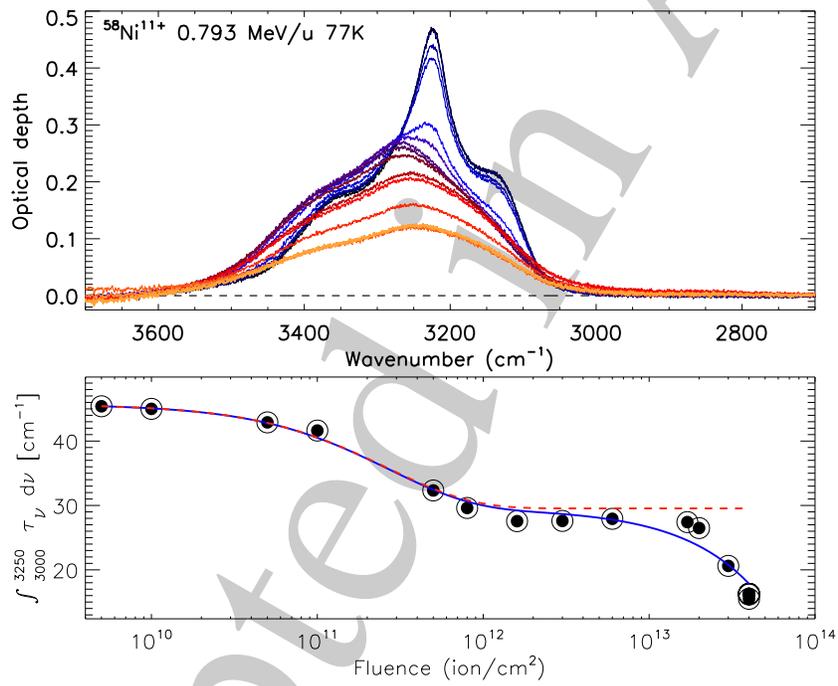}
\caption{Same as Fig.~\ref{Ni_13_6} for 77~K. See the experiments section for the difference in deposition history for this sample.}
\label{Ni_77}
\end{center}
\end{figure}

\begin{figure}[htbp]
\begin{center}
\includegraphics[height=\columnwidth,angle=90]{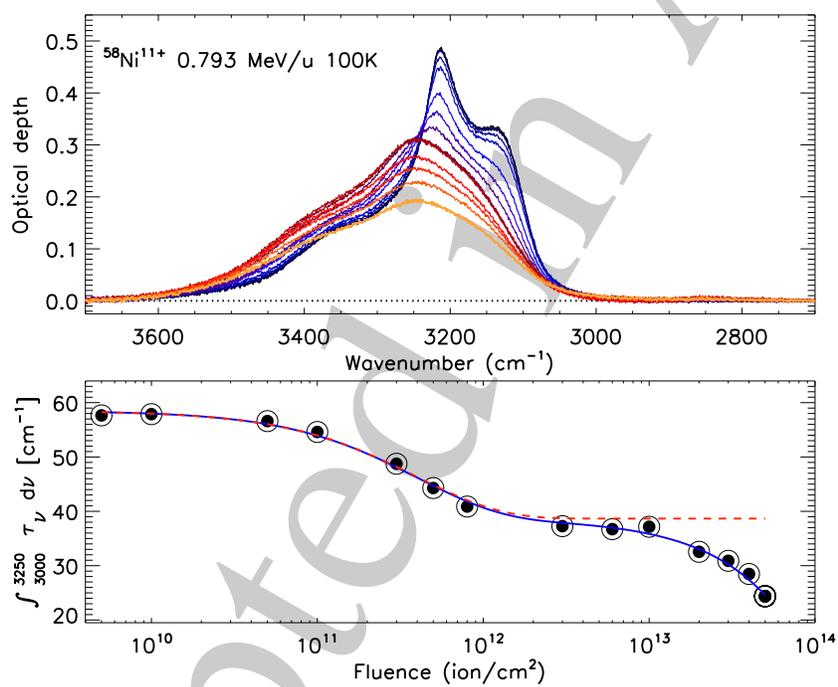}
\caption{Same as Fig.~\ref{Ni_13_6} for 100~K.}
\label{Ni_100}
\end{center}
\end{figure}

\begin{figure}[htbp]
\begin{center}
\includegraphics[height=\columnwidth,angle=90]{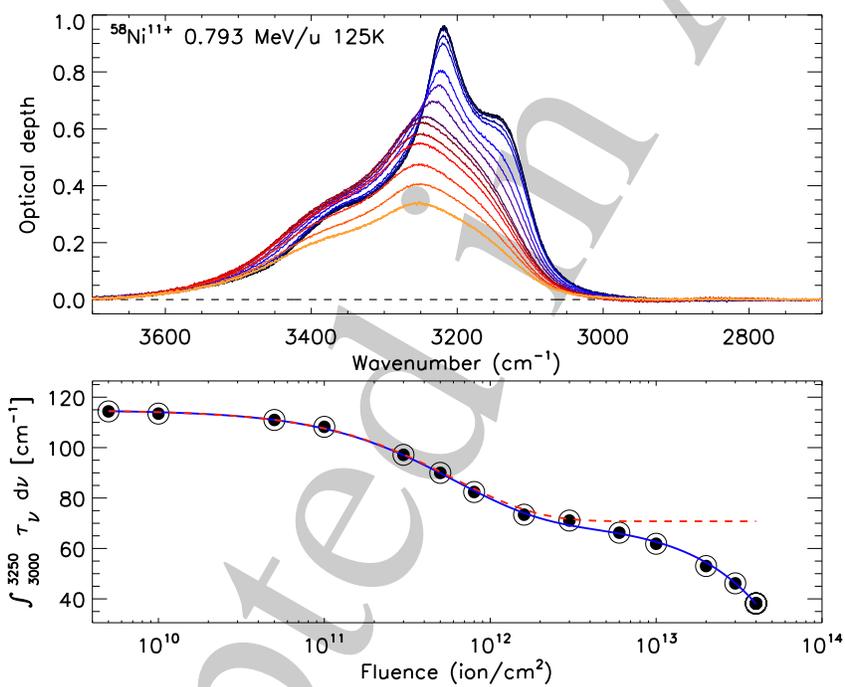}
\caption{Same as Fig.~\ref{Ni_13_6} for 125~K.}
\label{Ni_125}
\end{center}
\end{figure}

\begin{figure}[htbp]
\begin{center}
\includegraphics[height=\columnwidth,angle=90]{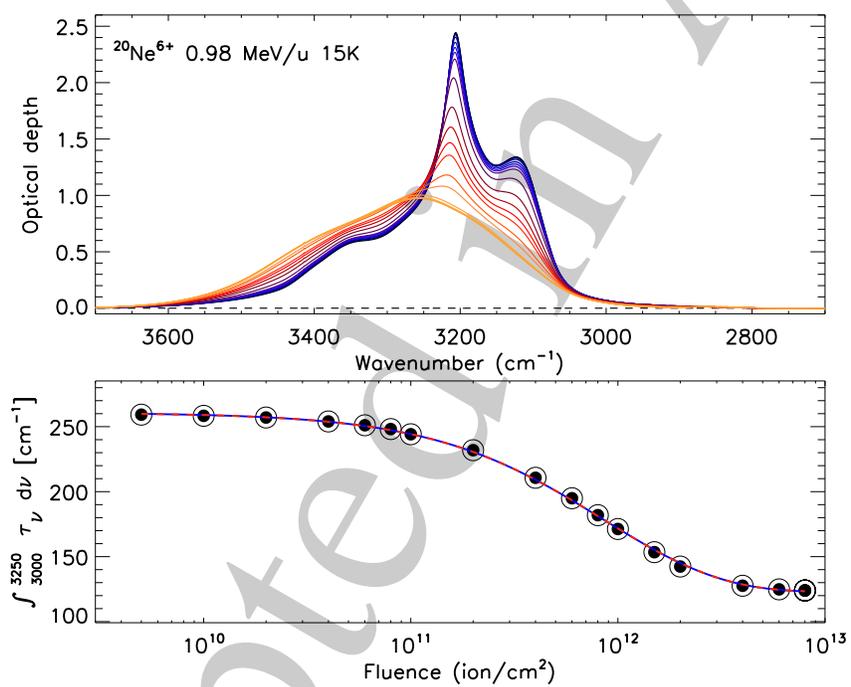}
\caption{Same as Fig.~\ref{Ni_13_6} for 19.6 MeV Ne swift ion irradiation at 15~K.}
\label{Ne_15}
\end{center}
\end{figure}

\begin{figure}[htbp]
\begin{center}
\includegraphics[height=\columnwidth,angle=90]{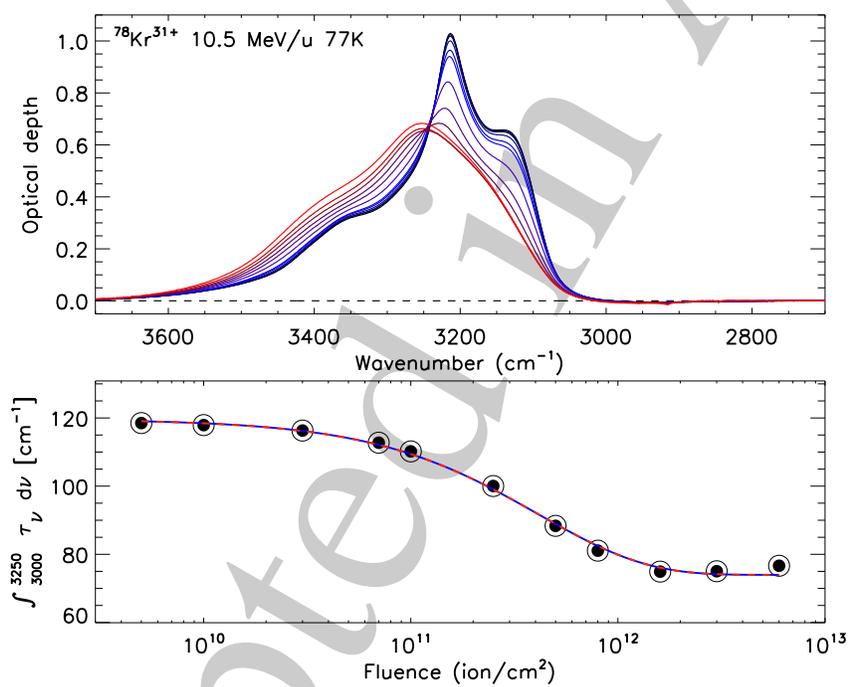}
\caption{Same as Fig.~\ref{Ni_13_6} for 819 MeV Kr swift ion irradiation at 77~K.}
\label{Kr_77}
\end{center}
\end{figure}

\begin{figure}[htbp]
\begin{center}
\includegraphics[height=\columnwidth,angle=90]{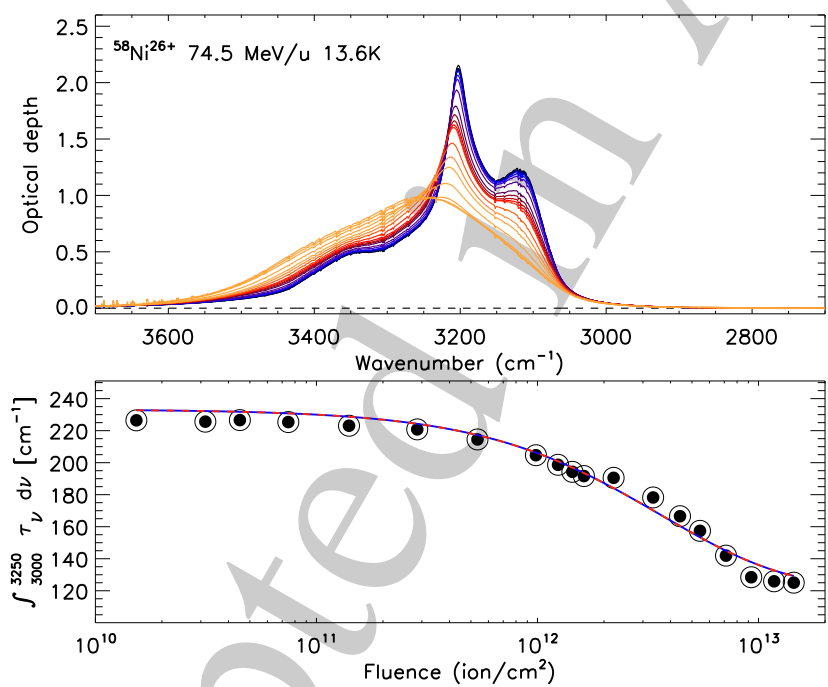}
\caption{Same as Fig.~\ref{Ni_13_6} for 3.64 GeV Ni swift ion irradiation at 13.6~K.}
\label{Ni_15}
\end{center}
\end{figure}

\begin{figure}[htbp]
\begin{center}
\includegraphics[height=\columnwidth,angle=90]{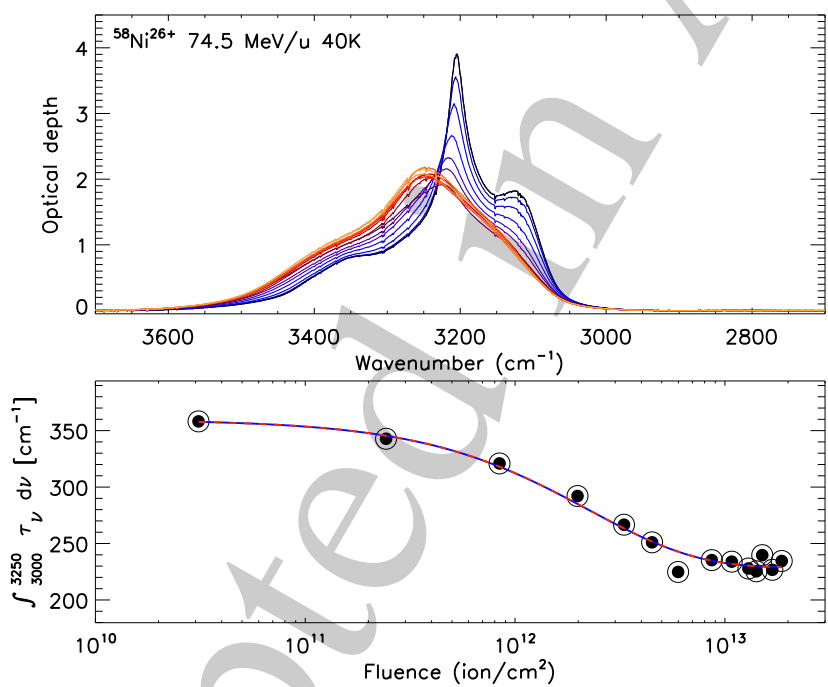}
\caption{Same as Fig.~\ref{Ni_13_6} for 3.64 GeV Ni swift ion irradiation at 40~K.}
\label{Ni_40}
\end{center}
\end{figure}

\begin{figure}[htbp]
\begin{center}
\includegraphics[height=\columnwidth,angle=90]{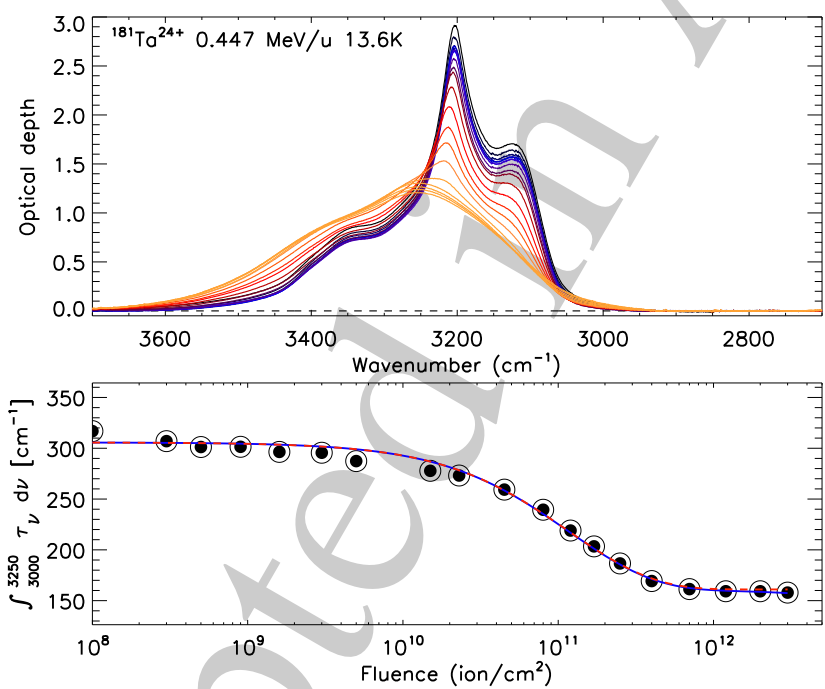}
\caption{Same as Fig.~\ref{Ni_13_6} for 81 MeV Ta swift ion irradiation at 13.6~K}
\label{Ta_13_6}
\end{center}
\end{figure}

The measured, baseline-corrected, optical depth spectra evolution, in the water ice stretching mode spectral region, upon irradiation, are displayed in the upper panels of Figs. \ref{Ni_13_6} to \ref{Ta_13_6} for different temperatures and different ions. The corresponding integrated optical depth $\rm I_{OD}$ as a function of fluence is shown in the lower panels. The model fit is overplotted, and the fit with the amorphisation cross section without the sputtering contribution is also shown. The derived cross-sections are given in Table~\ref{summary}, together with the sputtering yield estimate when possible.

%
\begin{landscape}
\begin{table*}[htdp]
\caption{Summary of experiments: E - incident ion energy; Se - electronic stopping power; Sn - nuclear stopping power; T - ice temperature; $\sigma$ - compact amorphisation cross-section; K$^{-1}$ - compact amorphisation dose; Ys - sputtering yield}
\begin{center}
\begin{tabular}{l l l l l l l l l l l l}
\hline
Ion                     &E  &Se$^a$     &Sn/Se  &Ice T &Ice film &Projected &$\sigma$     &1/K$\, ^d$ &Y$\rm_s$ &Reference\\
                        &   & & &       &thickness$^b$ &range$^c$ &     &       &       &/exp.\\
                        &(MeV)   &(eV/$\AA$) &(x10$^{-3}$)      &(K) &($\mu$m) &($\mu$m) &(10$^{4}$ $\AA^2$)     &(eV/H$_2$O)  &(H$_2$O/ion)   &date\\
\hline
Ni$^{26+}$      &3640                   &62.85  &0.42   &13.6   &0.39           &3.4 x10$^{3}$&0.16$\pm$0.12      &1.25$^{+3.75}_{-0.54}$ &-$\,^e$ &27-03-2014\\
                        &               &       &       &40   &0.62 &3.4 x10$^{3}$&0.28$\pm$0.14      &0.72$^{+0.75}_{-0.24}$ &-$\,^e$ &27-03-2014\\
\hline
Ne$^{6+}$               &19.6                   &143.1  &1.24   &15  &0.45      &16.1                 &0.62$\pm$0.09      &0.74$^{+0.12}_{-0.1}$ &-$\,^e$ & 09-06-2012\\
\hline
Kr$^{31+}$              &819            &409.3  &0.65   &77  &0.22      &175                 &1.65$\pm$0.26      &0.79$^{+0.15}_{-0.11}$ &-$\,^e$ & 11-11-2014\\
\hline
Ni$^{11+}$      &46.0   &461.5  &2.8    &13.6   &0.10   &19.4 &1.9$\pm$1.2              &0.78$^{+1.43}_{-0.3}$ &2610$\pm$1500&{03-05-2014}\\
                        &               &               &       &77     &0.10   &19.4 &2.65$\pm$2.0   &0.56$^{+1.72}_{-0.24}$ &3060$\pm$1690&{02-05-2014}\\
                                &               &               &       &100 &0.11 &19.4 &1.8$\pm$0.73 &0.82$^{+0.57}_{-0.24}$ &3200$\pm$1080&{03-05-2014}\\
                        &               &               &       &125    &0.23   &19.4 &1.14$\pm$0.23  &1.30$^{+0.33}_{-0.22}$  &9100$\pm$2400 &{03-05-2014}\\
\hline
Ta$^{24+}$      &81     &793.5  &15.9   &13.6  &0.58    &20.3   &5.02$\pm$2.97    &0.51$^{+0.73}_{-0.19}$ &17400$\pm$11800      & 23-09-2011\\
\hline
\end{tabular}
\end{center}
$^a$ The values are based on SRIM/TRIM calculations for pure ice. The density considered for the crystalline ice is 0.93g/cm$^3$ \citep[][]{Feistel2006}.
$^b$ approximate thickness assuming a band strength value of 2.7$\times$10$^{-16}$cm/molecule for the crystalline-ice stretching mode.
$^c$ calculated for a hypothetical semi-infinite ice target.
$^d$ K$^{-1}$ = Se / ($\sigma$ n), in eV/molecule
$^e$ insufficient signal-to-noise or fluence to provide a meaningful number
\label{summary}
\end{table*}
\end{landscape}
%

\section{Discussion}
\label{discussion}

\subsection{Amorphisation}
\label{sec:amorph}

\begin{figure}[htbp]
\begin{center}
\includegraphics[height=\columnwidth,angle=90]{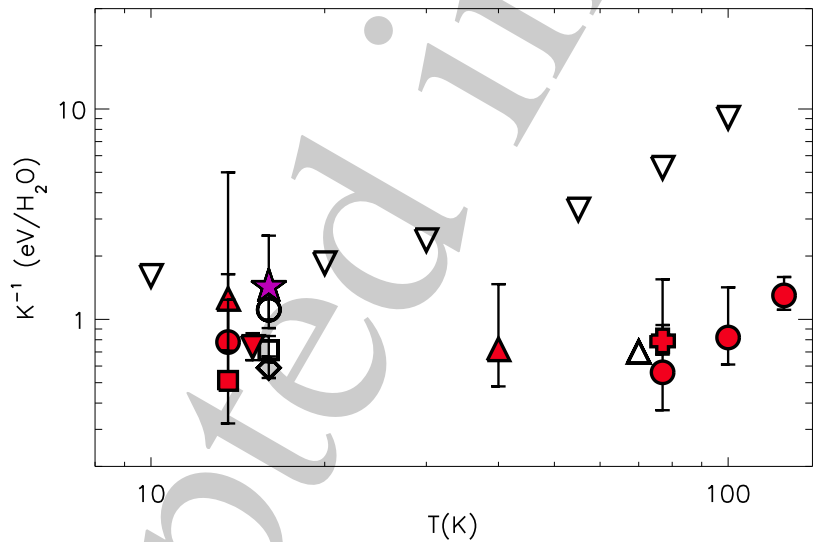}
\caption{K$^{-1}$ amorphisation dose as a function of crystalline ice temperature for the present measurements (red filled symbols: triangle, Ni$^{26+}$; downward triangle, Ne$^{6+}$; cross, Kr$^{33+}$; circle, Ni$^{11+}$; square, Ta$^{24+}$) and previous measurements following the evolution of the same ice feature \citep[open symbols: downward triangle, 3 keV He$^+$; diamond, 30 keV H$^+$; square, 30 keV He$^+$; circle, 60 keV Ar$^{2+}$; upward triangle, 100 keV Ar$^+$; see Table~1 in][]{Fama2010}. The purple star shows K$^{-1}$ for crystalline ice irradiated with FUV Lyman $\alpha$ photons, for comparison \citep{Leto2003}.}
\label{fig_K}
\end{center}
\end{figure}

The radiation-induced amorphisation of crystalline ice has been summarised recently for many experiments in \cite{Fama2010}, mainly for light ions (except Ar) and electrons in the 3-800 keV range. 
The evolution of the compact amorphous phase was estimated as a function of irradiation dose D (energy deposited per molecule), following an exponential behavior:
\begin{equation}
\rm w_a= 1-exp(-K_{cryst}D) = 1-w_c=1-exp(-\sigma_{am}F)
,\end{equation}
where
\begin{equation}
\rm D[eV/molecule]=\frac{F [ions/cm^2]\times S_{total}[eV/cm]}{n[molecule/cm^3]}
;\end{equation}
then
\begin{equation}
\rm K_{cryst}^{-1}[eV/molecule]= S_{total} / (\sigma_{am} \times n)
.\end{equation}
For simplicity, unless otherwise stated, it is written as $\rm K^{-1}(=~K_{cryst}^{-1})$.
In Fig.~\ref{fig_K} we display the amorphisation dose K$^{-1}$ (in eV/molecule) using our derived $\sigma_{am}$, compared to previously reported data using the same infrared stretching-mode analysis.\\
\cite{Fama2010} point out similarities and discrepancies between results obtained in different laboratories, also using various methods to follow
the amorphisation process (IR spectroscopy, electron diffraction, and channelling). Part of these discrepancies are probably attributable to different preparations for the crystalline ice film (annealed, deposited at high temperature and lowered to the temperature of irradiation), as well as various simplifications for the fitting method to the amorphisation fraction, including a maximum fraction of the amorphised ice or not doing so.\\
As reported, except for the \cite{Strazzulla1992} measurements, these experiments yield full amorphisation at high fluences. Our measurements for swift heavy ions are in very good agreement with those for light ions at the lowest temperature (13-16~K) and with the Ar irradiation at higher temperature. K$^{-1}$ increases with the temperature, showing an increasing resistance to amorphisation with temperature as one approaches the amorphous to crystalline transition, but swift heavy ions seem more efficient at higher temperatures than do light ions.\\
The ion-induced amorphisation of crystalline ice reported in the keV to hundreds of keV with electrons and light ions \citep[see][summary]{Fama2010} shows that an increasing amount of deposited energy is needed to amorphise the ice when the temperature increases. 
The heavier Ar ions \citep[100~keV Ar$^+$;][]{Baragiola2005} do not follow this trend, which is in line with our results for swift heavy ions, but including a significant nuclear stopping-power contribution
for Ar.\\
Comparing the measured K$^{-1}$ in the 70-80~K temperature range, derived for 800~keV protons \citep[K$^{-1}$$\approx$700, far infrared ice bands, ][]{Moore1992}, 100~keV electrons \citep[K$^{-1}$$\approx$10,  diffraction,][]{Lepault1983}, 100~keV Ar$^+$ \citep[K$^{-1}$$=$0.7, infrared band, ][]{Baragiola2005}, and this work (819 MeV Kr$^{33+}$, K$^{-1}$$=$0.79$^{+0.15}_{-0.11}$ and 46 MeV Ni$^{11+}$, K$^{-1}$$=$0.56$^{+1.72}_{-0.24}$), the cross-section dependence varies by orders of magnitude and suggests a more complicated relation than the one related only to the dose, at least for electrons, light ions, and moderate temperatures. 
K$^{-1}$ even diverges for electron irradiations above 70~K. Such a behaviour also seems true for photons \citep{Kouchi1990}, and it has been suggested that the presence of a mechanism that restores the original structure explains the strong temperature dependence of electrons and photons. 
One may speculate that defects in the crystalline ice network, induced by the irradiation, may be more efficiently annealed for the lower energy densities deposited by light ions and electrons, because the restoration would involve isolated water molecules.
This would explain the strong temperature dependence of K$^{-1}$.
This mechanism may become much less efficient for high-energy density deposition, such as in the heavy ion case, that produces individual tracks of amorphized ice.
The investigation of these differences should be further explored, because although they are less abundant in cosmic rays, if they possess such a flatter temperature dependence,
the heavy ions should play a more important role at intermediate temperatures for the crystalline ice amorphisation.
%


For temperatures below 100~K, the average crystalline ice K$\rm _{cryst}^{-1}$ amorphisation\footnote{crystalline to amorphous compact ice} dose that we deduce is about $0.76\pm0.24$ eV/molecule. The corresponding low-temperature porous amorphous ice compaction K$\rm_{amorph}^{-1}$ dose required is $0.24^{+0.24}_{-0.1}$ eV/molecule \citep[][]{Dartois2013}, i.e. about three times lower. The corresponding astrophysical time scales to evolve from the crystalline to the compact amorphous phase are thus three times larger than the ones given for low temperature porous amorphous ice compaction in Table 3 of \cite{Dartois2013}.

\subsection{Sputtering}
\label{sec:sputtering}

\begin{figure}[htbp]
\begin{center}
\includegraphics[height=\columnwidth,angle=90]{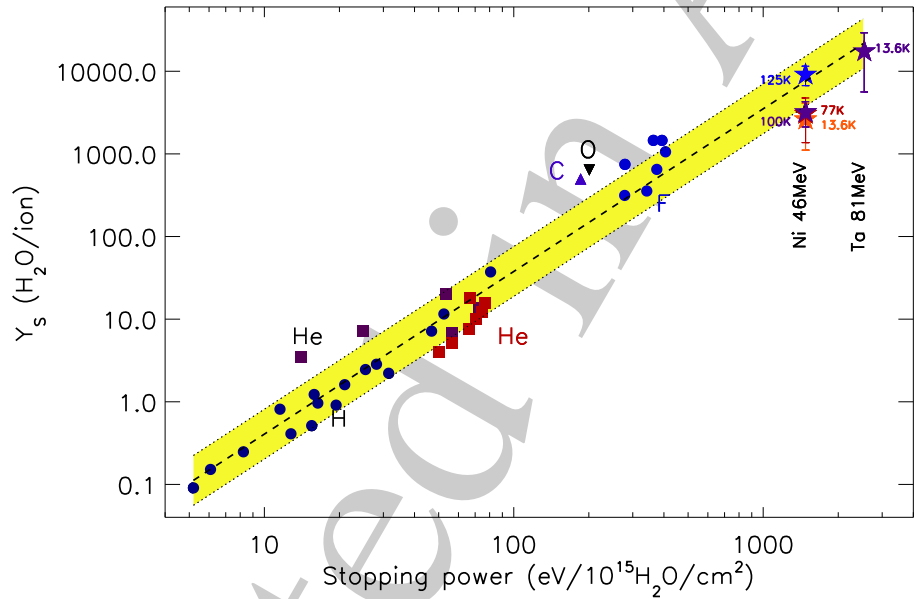}
\caption{Water-ice sputtering yield as a function of electronic stopping power. The result of our measurements (data above $\rm 1000 eV/10^{15} H_2O /cm^2$) are displayed along with the data compiled in Brown et al. (1984, see their Fig. 5 for details on the compiled data) at lower stopping power. The fitted dependence of the yield is given by the dashed line, and the coloured region shows the standard error of estimate on the yield. See text for details.}
\label{fig_sputtering}
\end{center}
\end{figure}

In the ion irradiations performed at high electronic stopping power (Ni$^{11+}$ and Ta$^{24+}$), the sputtering of ice can be estimated from the infrared spectrum at high fluence. This yield is displayed in Fig.~\ref{fig_sputtering}, together with previous measurements reported by \cite{Brown1984}.\\
The data display a power law dependence, which can well be fitted with
\begin{equation}
\rm Yield[H_2O/ion] = \alpha S_e^{\beta}[eV/10^{15}H_2O/cm^2]
\end{equation}
\begin{equation}
\rm  log_{10}(\alpha)=-2.36\pm{0.3} ; \beta=1.97\pm{0.07}.
\end{equation}
The sputtering yields for water ice measured in our experiments are slightly below the yields extrapolated by taking only the data points from \cite{Brown1984} into account. 
This may arise from the differences in ice sample preparation or from the fact that the measurements via the infrared are only sensitive to the yield at large fluence, when the crystalline ice has already been amorphized (into a compact amorphous phase) and the first layers have already been sputtered. Part of the difference in the yield might thus arise from the ice already having been processed. Such a difference could be addressed with direct desorption measurements in the same swift ion range with the very first ions impinging the target, using another detection technique.

\subsection{Ion sputtering versus photodesorption}

The sputtering yield dependence can be used to infer the ion-induced sputtering of H$_2$O ice in a dense cloud. 
This is evaluated by integrating the product of this cross section with the cosmic ray flux over their energy and abundance distribution.
The detail of the calculation can be found for the low-temperature porous amorphous ice compaction in \cite{Dartois2013}, so we refer the reader to this article. 
For the differential cosmic ray flux, we use the same functional form. The form parameter 
$\rm E_0 (MeV)$ will affect the differential flux of low-energy cosmic rays, with no incidence on the high energy.
We adopt the galactic cosmic rays (GCR) distribution of cosmic ray abundances. In the equations 
we replace the low-temperature, porous amorphous-ice-compaction cross-section by the sputtering yield derived above and derive the H$_2$O ice sputtering rates reported in Table~\ref{rates}.

%
%
\begin{table}[htdp]
\caption{H$_2$O-ice sputtering rates}
\begin{center}
\ra{1.3}
\begin{tabular}{@{}rrrrcrrr@{}}
\toprule
& \multicolumn{3}{c}{$\rm GCR$} \\ 
\cmidrule{2-4} 
$\rm E_0 (MeV)$ & $200$ & $400$ & $600$ \\ 
\cmidrule{2-4}%
$\rm \zeta (s^{-1}) ^a$         &3.34(-16)      &5.89(-17)      &2.12(-17)      \\ 

$\rm \eta^{GCR} (H_2O/cm^{2}/s) ^b$             &27.0$^{+26}_{-13}$             &8.3$^{+8.2}_{-4.1}$            &4.1$^{+4.1}_{-2.1}$            \\ 
 \bottomrule
\end{tabular}
\end{center}
$^a$ calculated ionisation rate; 
$^b$ H$_2$O sputtering rate.
\label{rates}
\end{table}

In the shielded part of a dense cloud, the cosmic rays induced for UV photon flux is FUV$\approx$10${^4}$ photons/cm$^2$/s \citep[e.g.][]{Shen2004}.
Therefore, the direct cosmic ray water-ice sputtering will contribute equally to a FUV photodesorption yield $\rm\eta^{FUV}$ for
\begin{equation}
\rm \eta^{FUV} = \eta^{GCR}/FUV \approx 10^{-3},
\end{equation}
where $\rm\eta^{GCR}$ is the CR H$_2$O sputtering rate. The FUV water-ice photodesorption yield is still under debate (e.g., \cite{Kulikov2010},$<$0.05 molecule/photon; \cite{Oberg2009}, $\sim$1.6$\times$10$^{-3}$ molecule/photon at T=10~K; \cite{Westley1995}, $\sim$3-5$\times$10$^{-3}$ molecule/photon  for 35$\lsim$T$\lsim$85~K, \cite{Desimone2013} $\approx$7$\times$10$^{-5}$ molecule/photon at 100~K for 157~nm photons), because the wavelength-dependent absolute yields and their temperature evolution are more difficult to measure than species such as CO or N$_2$, which are less bonded to the ice surface.\\
%
In dense clouds, which are shielded from the external interstellar standard radiation field (ISRF), the FUV photons are induced by the impact excitation of molecular hydrogen by secondary electrons generated by cosmic rays \citep[e.g.][]{Gredel1989}. Therefore, the magnitude of the photodesorption yield and the direct sputtering rate by cosmic ray ions are related. From the fitted cross section to the sputtering yield measured data, it seems clear that the cosmic-ray-induced water-ice desorption competes with the FUV photodesorption yield. 




\section{Conclusion}

Swift heavy ion irradiation of crystalline water ice, at high energies, amorphise the ice
efficiently and release water molecules by surface sputtering. The main conclusions of our study follow.\\

- The experimentally measured swift ion dose (K$^{-1}$, eV/molecule) required to amorphise crystalline ice is about three times higher than the dose for the compaction of low-temperature porous amorphous ice, which suggests a better resilience to a phase change than the low-temperature porous amorphous ice.\\

- The measured temperature dependence of the cross-section for the crystalline to compact amorphous-ice transition shows a flatter dependence than measured for electrons and/or light ions. The differences should be investigated further, in particular to constrain to what extent the  deposited energy density, much higher with heavy ions, has an impact on the temperature dependence, in addition to the dose that is often used as the reference to compare cross-sections.\\

- The water ice sputtering yields, which were derived at large fluence irradiations when the ice is amorphized, are in good agreement with measurements at lower electronic stopping powers using a different analytical technique. The dependence of the sputtering yield is quadratic over three orders of magnitude of electronic stopping power.\\

- Combining, in a simple model, the galactic cosmic ray distribution of ion abundances and the measured water-ice-molecule sputtering yield, the H$_2$O-ice sputtering rate in dense clouds is estimated in the range $\rm \eta^{GCR}=$4.1$^{+4.1}_{-2.1}$ $\rm H_2O/cm^{2}/s$ at a relatively low ionisation rate $\rm \approx2\times10^{-17}s^{-1}$, and rises to 27.0$^{+26}_{-13}$ $\rm H_2O/cm^{2}/s$ at a higher ionisation rate $\rm \approx3\times10^{-16}s^{-1}$.\\

The magnitude of this sputtering mechanism provides a sputtered H$_2$O-ice molecules major feedback on the gas phase, competing with the FUV photodesorption. They are nevertheless related, because FUV photons are mainly cosmic-ray-induced in the dense regions where water ice is observed.


\section*{Acknowledgements}
The authors would like to cordially thank the anonymous referee and the editor Malcolm Walmsley, for the comments that improved the scientific content of the article, as well as J. Adams for suggestions on the language and structure of the manuscript.
 The experiments were performed at the Grand Acc\'el\'erateur National d'Ions Lourds (GANIL) Caen, France. We would in particular like to thank the LISE staff. 
Part of this work was supported by the ANR IGLIAS project, grant ANR-13-BS05-0004 of the French Agence Nationale de la Recherche. 


%
\end{document}